\documentclass[prl,twocolumn,superscriptaddress,showpacs,floatfix]{revtex4}

\usepackage{epsfig,amsmath,amssymb,latexsym}

\usepackage{subfigure}

\def\afs{$A_y$Fe$_{1.6}$Se$_2$}
\def\kfs122{K$_y$Fe$_2$Se$_2$}
\def\kfsx22{K$_{1-x}$Fe$_2$Se$_2$}

\def\bfa122{BaFe$_2$As$_2$}
\def\fs11{FeSe}
\def\afsxy{$A_y$Fe$_{2-x}$Se$_2$}

\begin{document}

\title{Block Spin Magnetic Phase Transition of
A$_y$Fe$_{1.6}$Se$_2$ Under High Pressure} \preprint{1}

\author{Chao Cao}
 \email[E-mail address: ]{ccao@hznu.edu.cn}
 \affiliation{Condensed Matter Group,
  Department of Physics, Hangzhou Normal University, Hangzhou 310036, China}
\author{Minghu Fang}
  \affiliation{Department of Physics,
  Zhejiang University, Hangzhou 310027, China}

\author{Jianhui Dai}
\email[E-mail address: ]{daijh@zju.edu.cn}

\affiliation{Condensed Matter Group,
  Department of Physics, Hangzhou Normal University, Hangzhou 310036, China}

\affiliation{Department of Physics, Zhejiang University, Hangzhou
310027, China}

\date{August 12, 2011}

\begin{abstract}
We predict an unconventional magnetic ground state in \afs\ with
$\sqrt{5}\times\sqrt{5}$ Fe-vacancy superstructure  under hydraulic
external pressure based on first-principles simulations. While the
Fe-vacancy ordering persists up to at least $\sim $ 12GPa, the
magnetic ground state goes at $\sim$10GPa from the BS-AFM phase to a
N{\'e}el-FM phase, a ferromagnetic arrangement of a "{\it{N{\'e}el
cluster}}". The new magnetic phase is metallic, and the BS-AFM to
N{\'e}el-FM phase transition is accompanied by a sizable structural
change. The two distinct magnetic phases can be understood within
the extended $J_1$-$J_2$ Heisenberg model by assuming a
pressure-tuned competition between the intrablock and interblock
nearest-neighbor couplings of iron moments.
\end{abstract}

\pacs{75.25.-j,71.20.-b,75.10.Hk}
\maketitle

\begin{figure}[ht]
 \centering
 \subfigure[]{
   \scalebox{0.36}{\includegraphics{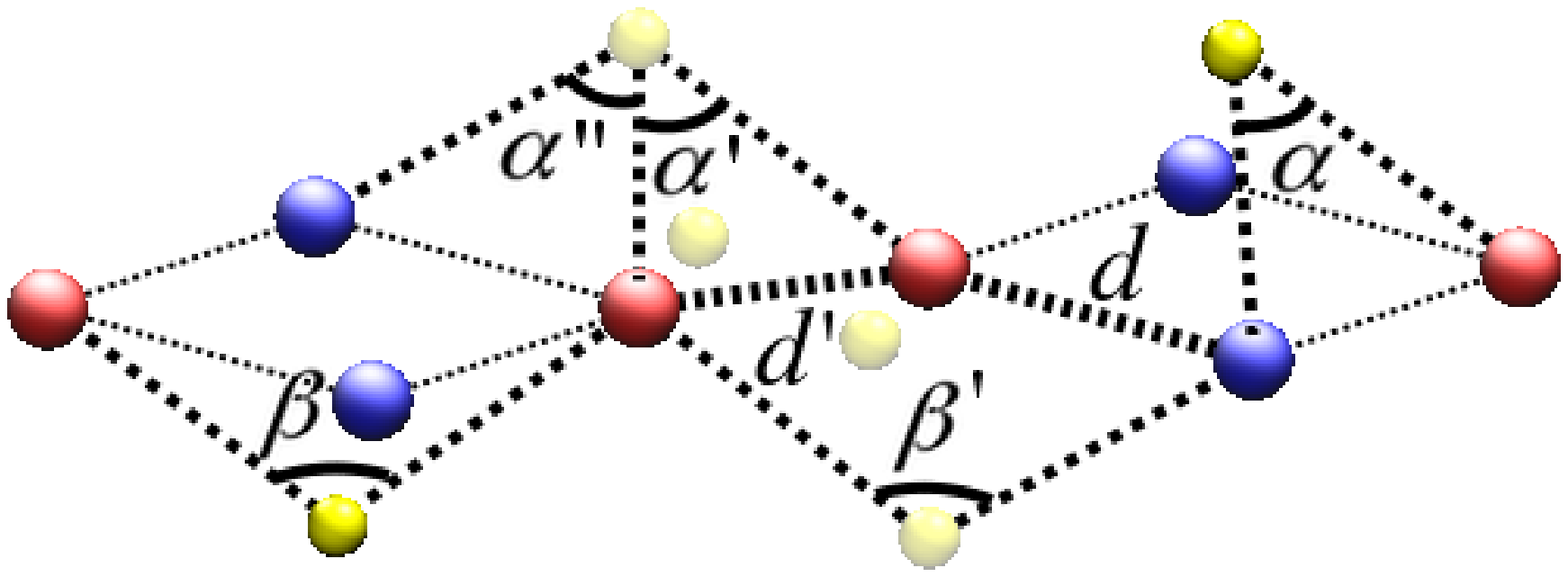}}
   \label{fig_para}
 }
 \subfigure[BS-AFM]{
   \scalebox{0.2}{\includegraphics{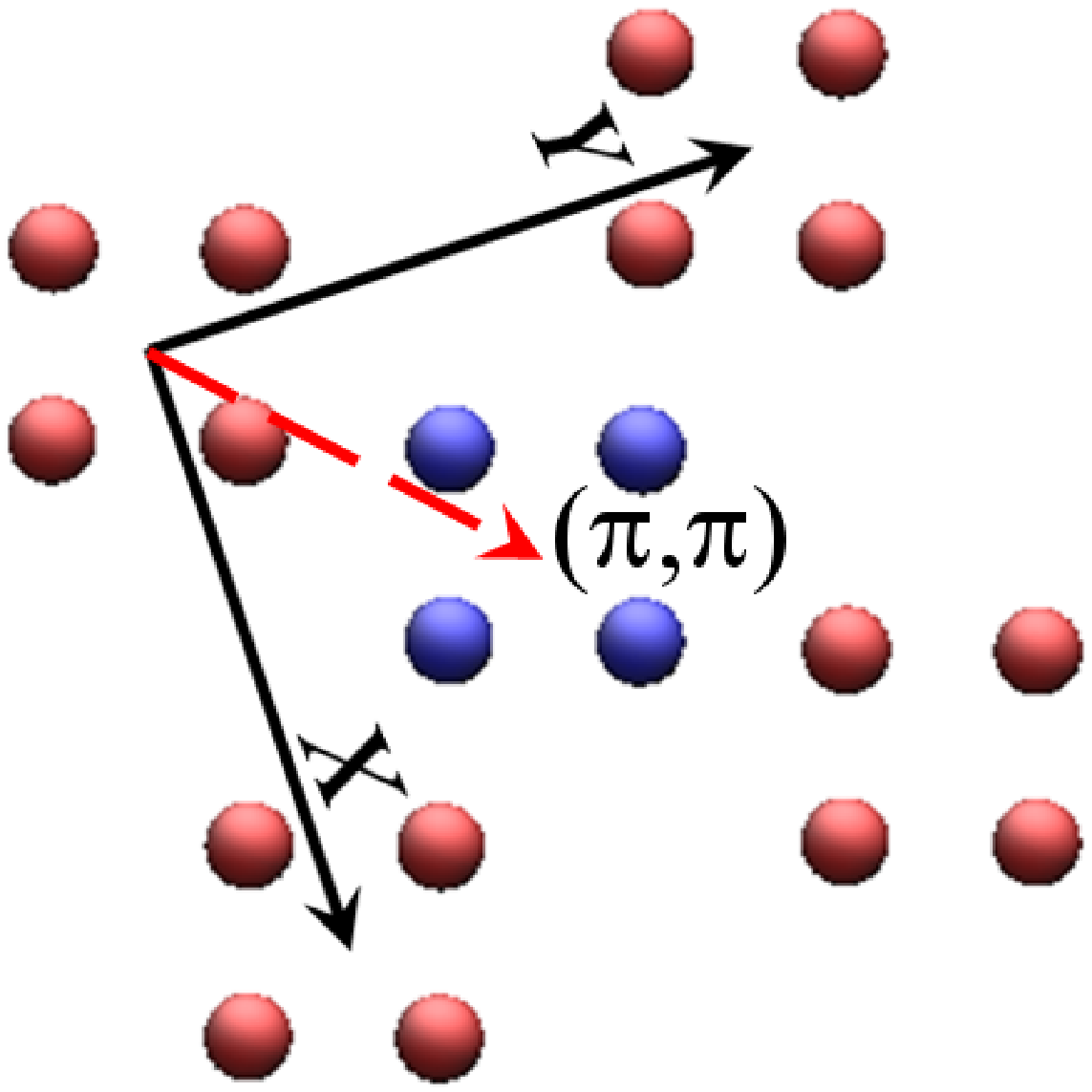}}
   \label{fig_geo_afm2}
 }
 \subfigure[N\'{e}el-FM]{
   \scalebox{0.2}{\includegraphics{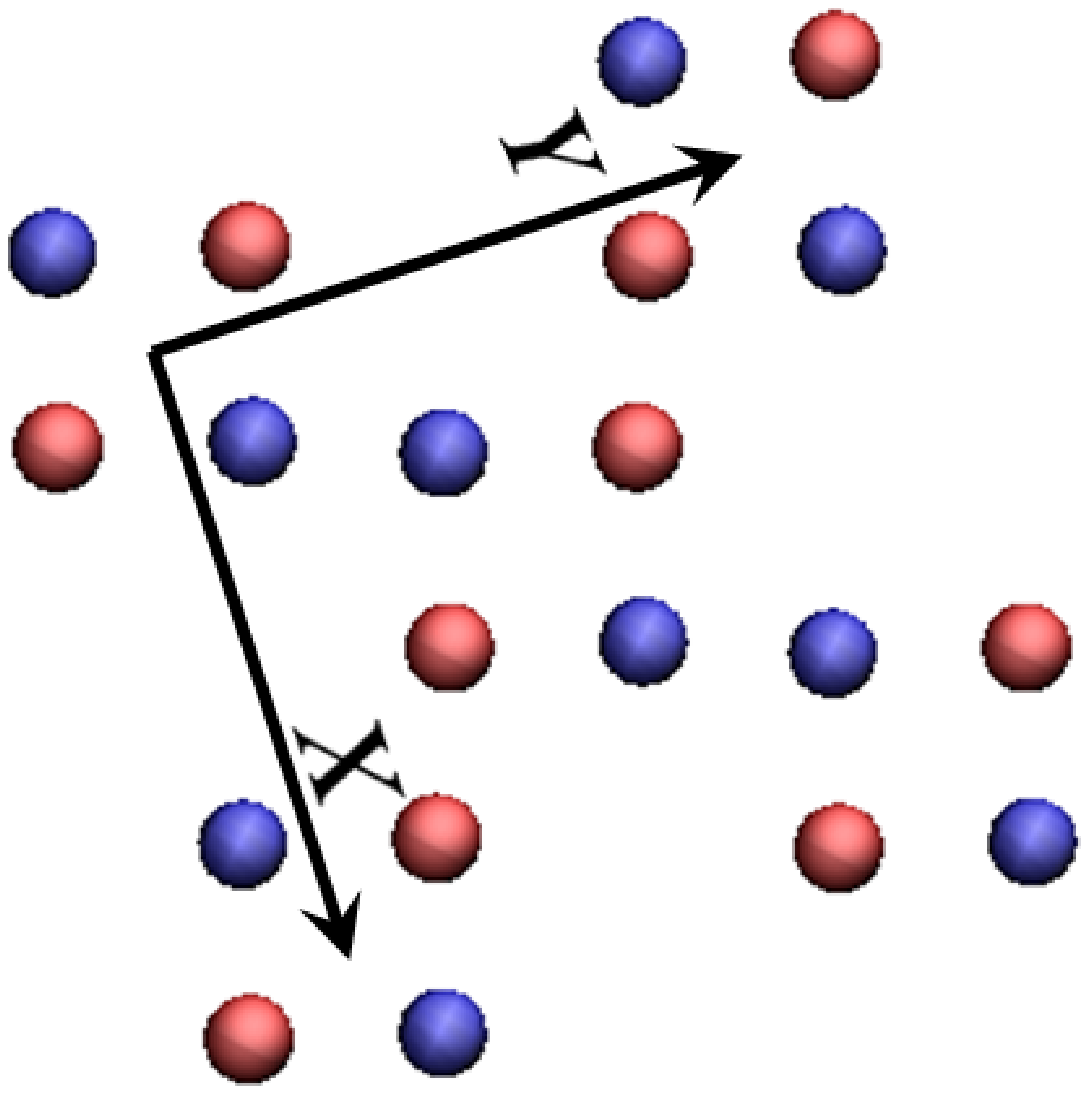}}
   \label{fig_geo_afm4}
 }
 \caption{The geometry (a) and magnetic patterns for (b) BS-AFM (AFM2) and (c) N\'{e}el-FM (AFM4). The structural parameters are as defined in FIG.\ref{fig_ev_pressure}.\label{fig_pattern}}
\end{figure}

{\it Introduction.} The anti-PbO-type FeSe is the simplest Fe-based superconductor with $T_c\sim$ 8K\cite{MKWu-11}. Recently, a class of new iron superconductors \afsxy\ ($A$=Tl,K,Rb,Cs) with enhanced $T_c\sim$ 30K\cite{xlchen_prb_82_180520,mhfang_1012,arxiv:1101.0462} has attracted intensive interest. These materials are structurally similar to the 122-type iron-pnictides\cite{Rotter-122} with the FeSe layers intercalated by the $A$ atoms, leaving certain amount of Fe-vacancies in the Fe-square lattice. The iron-vacancies are expected to order in some periodic superstructures at certain $x$ values\cite{mhfang_1012}. Among the proposed superstructures, the $\sqrt{5}\times\sqrt{5}$ vacancy ordering pattern(see in FIG. \ref{fig_pattern}), corresponding to $x=0.4$, seems to be of special importance since it exists in most of \afsxy\ compounds as confirmed by experiments from neutron diffraction\cite{arxiv:1102.0830,arxiv:1102.2882,C1SC00070E,0953-8984-23-15-156003} to high-resolution transmission electron microscope\cite{PhysRevB.83.140505}.

In addition to the $\sqrt{5}\times\sqrt{5}$ superstructure, a novel block-spin antiferromagnetism, the BS-AFM order shown in FIG.\ref{fig_pattern}(b) with large magnetic moment of irons, was discovered by the neutron diffraction experiments\cite{arxiv:1102.0830}. A remarkable observation is the co-existence of superconductivity and BS-AFM order in \afsxy\cite{arxiv:1102.0830}. It has been debated whether the co-existence is intrinsic at the microscopic level\cite{0295-5075-94-2-27008}, or due to phase separation\cite{arxiv:1106.3026}. Previous independent density functional theory (DFT) calculations show that the ground state has a BS-AFM order in the presence of the Fe-vacancy superstructure \cite{cao_dai_245,PhysRevB.83.233205}. Since the magnetic ordering temperature, which is unprecedentedly high ($T_N\sim 550 K$), is close to the Fe-vacancy ordering temperature $T_V\sim 580 K$\cite{arxiv:1102.0830}, it was also proposed that the Fe-vacancy ordering may be driven by magnetic exchange interactions so as to minimize the magnetic frustrations\cite{arxiv:1103.4599}.

All these interesting issues are closely related to the Fe-vacancy orderings and call for further experimental investigations. A key to resolve these issues is to clarify whether the BS-AFM is the only magnetic ground state or whether there are other magnetic ground states in the Fe-vacancy ordered compounds. Here we suggest to seek for these states by applying physical pressure, a clean parameter to tune the lattice and electronic structures. We note a recent high pressure experiment on superconducting sample of nominal K$_{0.8}$Fe$_{1.7}$Se$_{2}$ compound, where the resistance hump is suppressed at a critical pressure of 8.7 GPa and $T_c$ is suppressed at a similar pressure\cite{arxiv:1101.0092}. The metallic phase in the high pressure regime was believed to arise from charge transfer between two different Fe vacancy occupancies but its magnetic structure is unknown.

In this paper, we study the lattice and magnetic structures of \afs\ under high pressure up to 16 GPa by using first-principles simulations. We find that while the BS-AFM ground state (which is insulating at $y=0.8$\cite{cao_dai_245,PhysRevB.83.233205}) persists to 10GPa, a novel metallic magnetic phase, the N{\'e}el-FM phase, becomes the new ground state for higher pressures. Both the BS-AFM and N{\'e}el-FM phases can be described by the extended $J_1$-$J_2$ model while the magnetic phase transition is accompanied by a sizable structural change.

\begin{figure*}[ht]
 \centering
 \subfigure[]{
   \rotatebox{270}{\scalebox{0.4}{\includegraphics{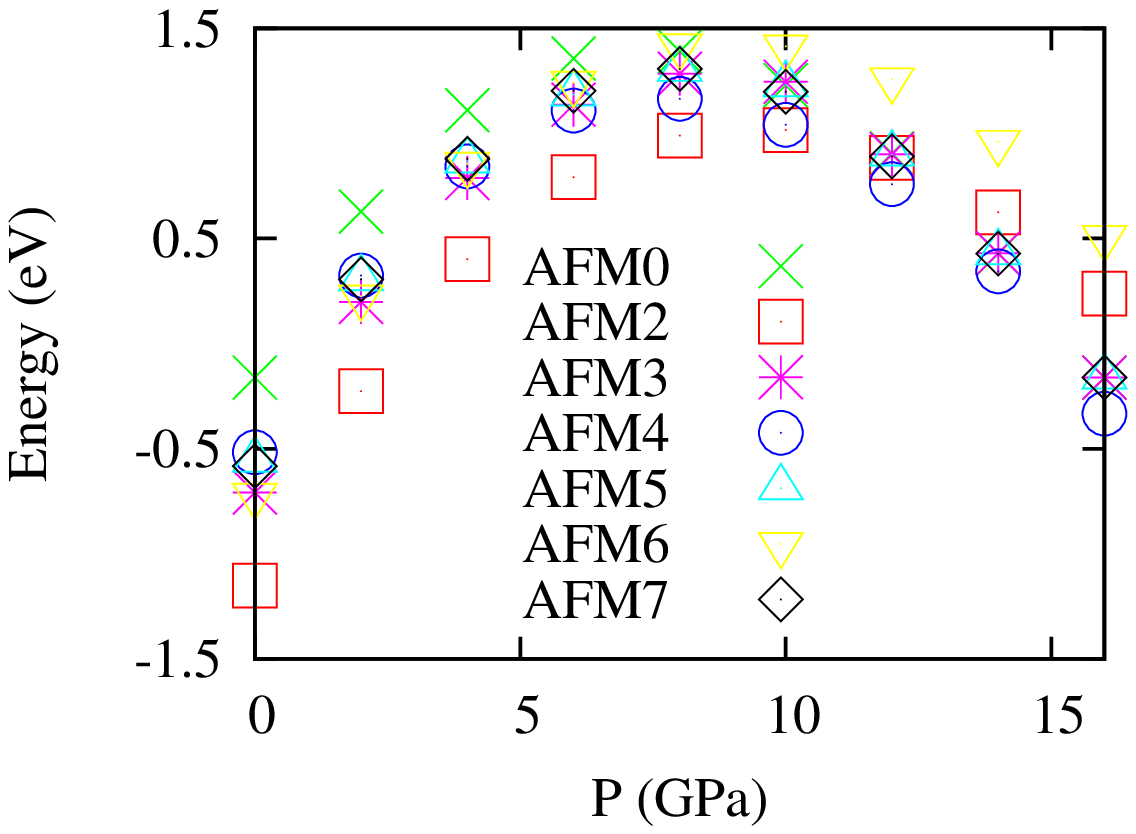}}}
   \label{fig_energy_pressure}
 }
 \subfigure[]{
   \rotatebox{270}{\scalebox{0.4}{\includegraphics{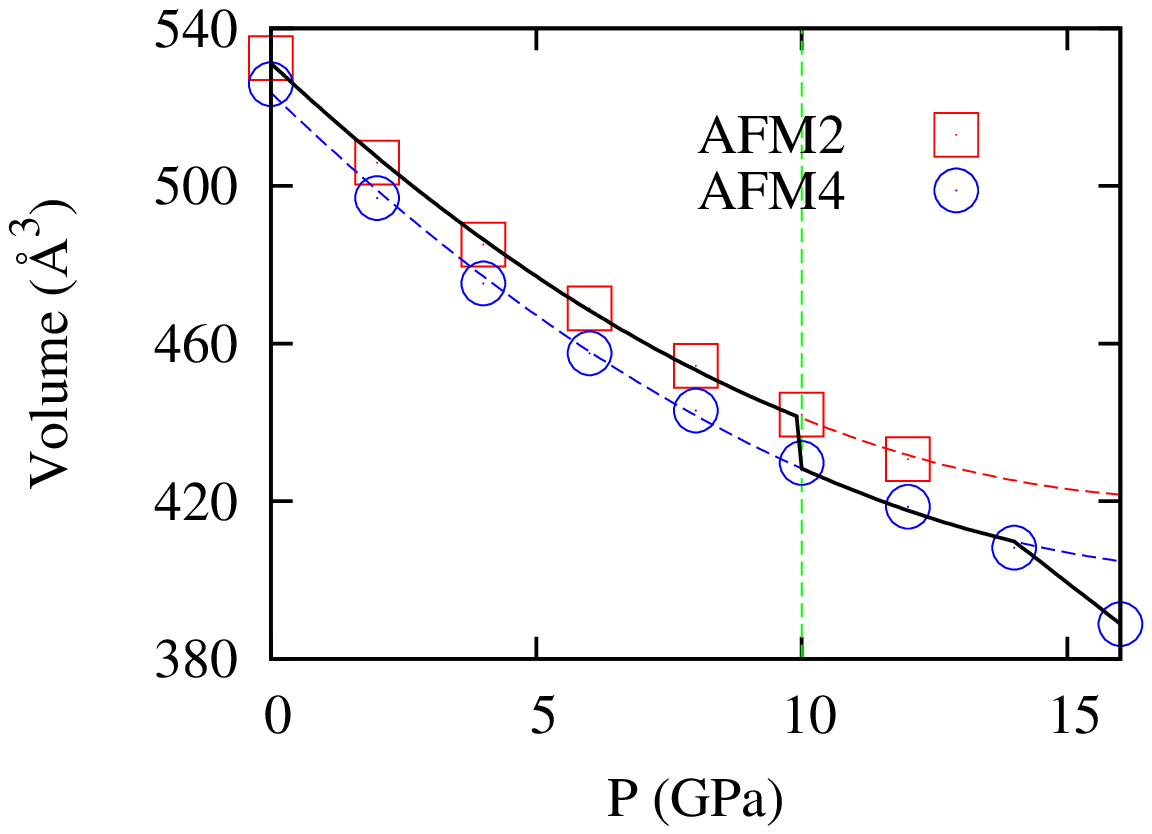}}}
   \label{fig_volume_pressure}
 }
 \newline
 \subfigure[$d_{\mathrm{Fe-Fe}}$]{
   \rotatebox{270}{\scalebox{0.4}{\includegraphics{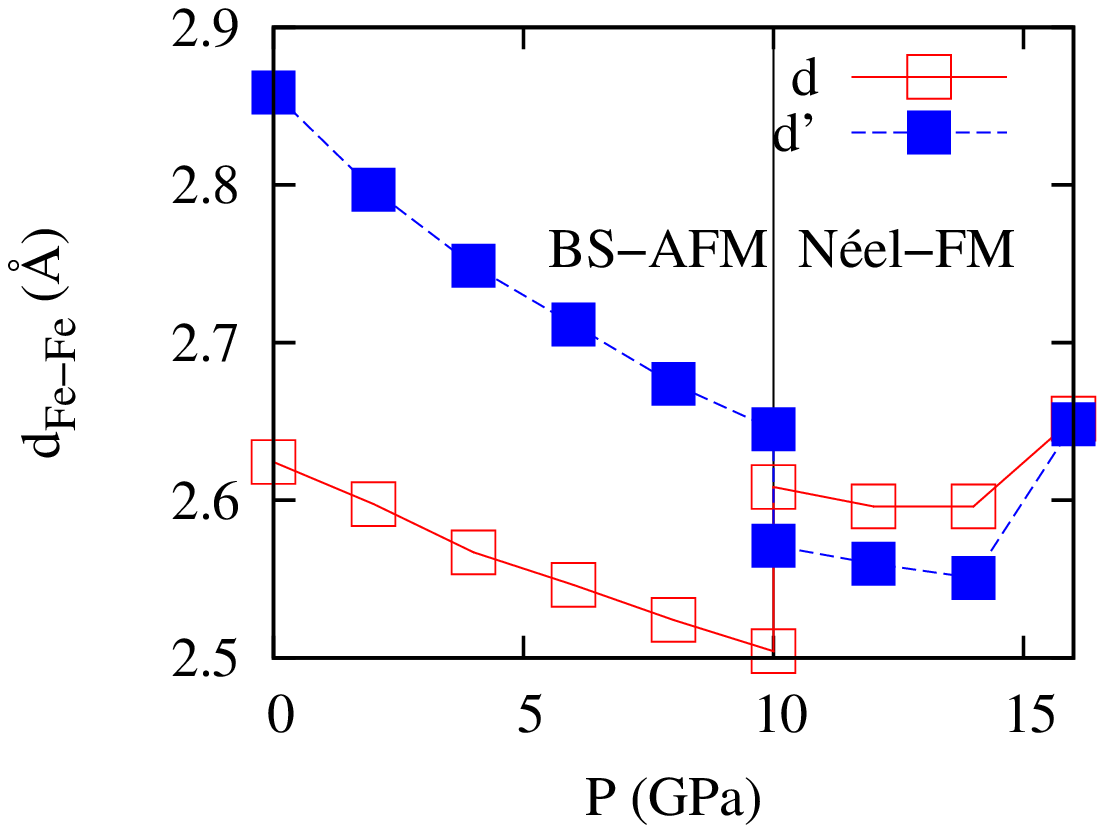}}}
   \label{fig_nn_length}
 }
 \subfigure[$\alpha_{\mathrm{Fe-Se-Fe}}$]{
   \rotatebox{270}{\scalebox{0.4}{\includegraphics{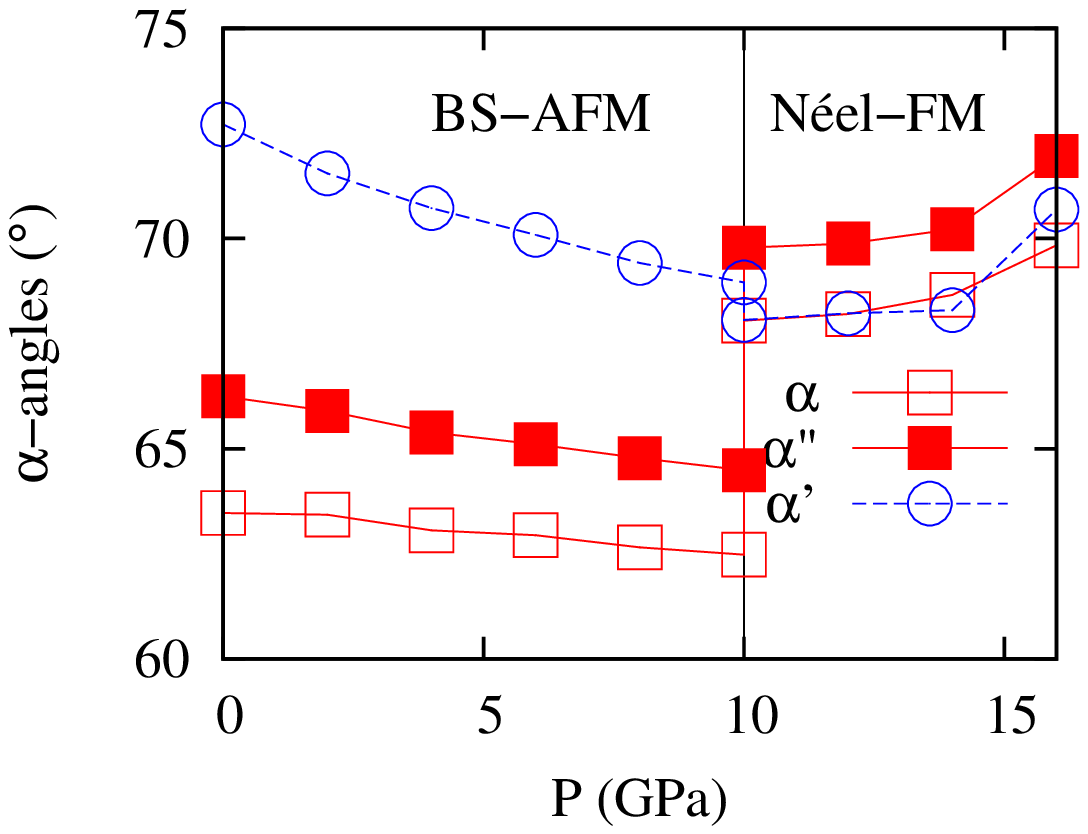}}}
   \label{fig_alpha_angle}
 }
 \subfigure[$\beta_{\mathrm{Fe-Se-Fe}}$]{
   \rotatebox{270}{\scalebox{0.4}{\includegraphics{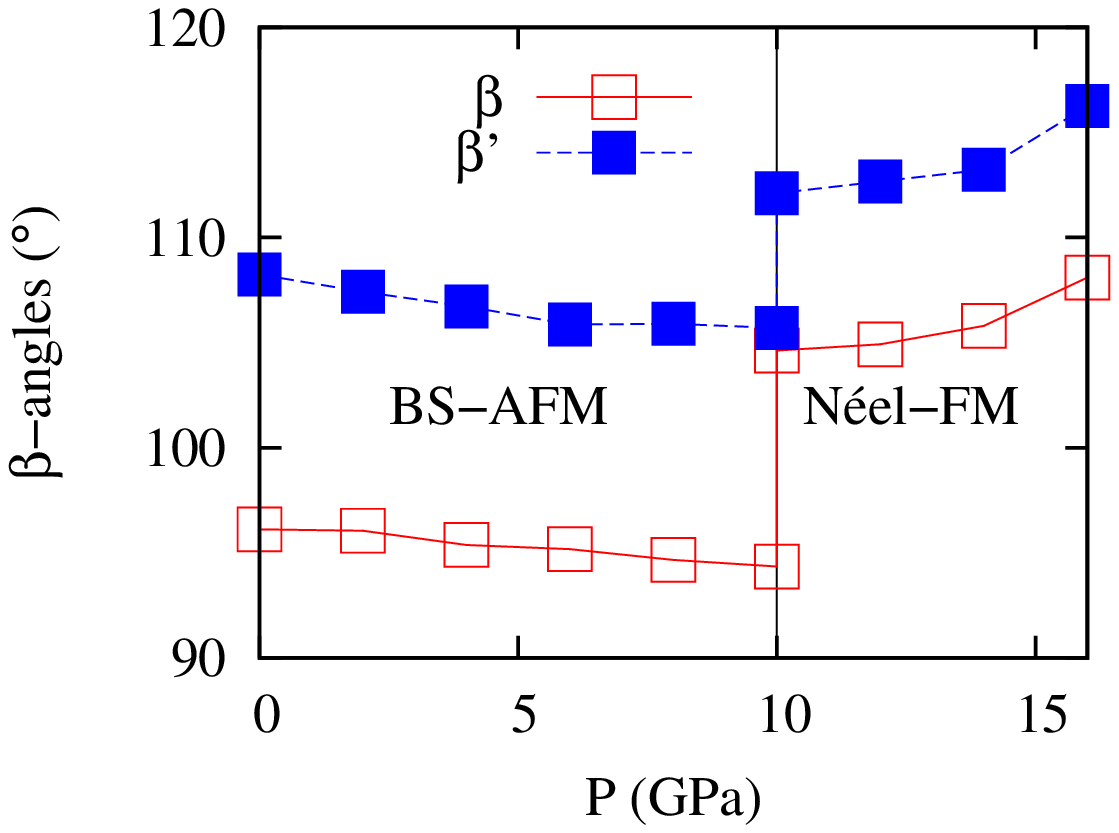}}}
   \label{fig_beta_angle}
 }
 \caption{Variation of magnetic configuration energies and structural parameters with respect to external pressure. (a) Total energy per unit cell of different magnetic phases. To enhance the visibility, the total energies have been renormalized using $\tilde{E}(P)=E(P)-kP+E_0$ with $k=2.800$ eV/GPa and $E_0=120.163$ eV. (b) Unit cell volume of BS-AFM and N\'{e}el-FM phases, where the black solid line indicates the actual volume per unit cell. (c)-(e) Various structural parameters as defined in Fig. \ref{fig_para}\label{fig_ev_pressure}}
\end{figure*}

{\it Magnetic phases and ground state.} For the first principle simulations, plane-wave basis and projected augmented wave methods are used as implemented in VASP code \cite{vasp_1,vasp_2}. The Perdew, Burke, and Ernzerhoff flavour of generalized gradient approximation (GGA)\cite{PBE_xc} is employed to calculate the electron exchange-correlation energy. A 360 eV energy cutoff for the plane-waves and $4\times4\times4$ Monkhorst-Pack k-grid\cite{mp_kgrid} are chosen to ensure the total energy converges to 1 meV/Fe. The external pressure is introduced using the Pulay stress method. With the applied pressure, both the internal coordinates and the lattice constants are fully optimized for each calculated magnetic and non-magnetic configurations until the total force on each atom is $<0.01$ eV/\AA. For the density of state (DOS) calculations, a dense $16\times16\times16$ $\Gamma$-centered k-grid, as well as the tetrahedra method are used to obtain accurate energy gap sizes. In the following text, we report and discuss detailed results for $A$=Tl, while the validity of all conclusions is checked for $A$=K and Rb.

We first identify the true ground state by examining the
relative energies of several possible magnetic
configurations\cite{cao_dai_245,note1}. As shown in FIG.
\ref{fig_ev_pressure}(a), the relative energetic order of different
magnetic configurations vary drastically with the applied external
pressure. The AFM2 pattern (the BS-AFM phase,
FIG.\ref{fig_pattern}(b)) remains the ground state until around
10GPa, where the originally second highest AFM4
(FIG.\ref{fig_pattern}(c)) pattern takes over and eventually becomes
the new ground state on the high pressure side. In the AFM4 phase,
the four spins in each Fe$_4$ block form a N\'{e}el order, while all
block configurations are parallelly aligned, thus we denote it the
N\'{e}el-FM phase for simplicity. The new ground state is at least
15 meV/Fe lower than all other calculated phases at 12 GPa, and is
at least 20 meV/Fe lower at 16 GPa.

{\it Structure variation.}  A close examination reveals the details
of the vacancy-induced structural distortion during the process, as
shown in FIG. \ref{fig_ev_pressure}(c)-(e). From 0 GPa to 10 GPa,
the interblock nearest-neighbor (n.n.) Fe-Fe distance
$d_{\mathrm{Fe-Fe}}'$ is compressed from 2.86\AA\ to 2.64\AA, more
severely than the intrablock n.n. Fe-Fe distance
$d_{\mathrm{Fe-Fe}}$ (from 2.62\AA\ to 2.50\AA). Meanwhile, all
Fe-Se-Fe angles ($\alpha$, $\alpha'$, $\alpha''$, $\beta$, and
$\beta'$) are slightly reduced. The formation of vacancy
superstructures also causes the relative height of Fe atoms
$z_{\mathrm{Fe}}$ (measured from the nearest Tl-layer) to be
slightly different in two neighboring blocks. However, this
distortion is negligible
($z^{1}_{\mathrm{Fe}}-z^{2}_{\mathrm{Fe}}<0.08$\AA) in the BS-AFM
state until 10 GPa.

The magnetic phase transition then occurs, together with a
structural change, identified by the first abrupt volume change
around 10GPa as shown in FIG.\ref{fig_ev_pressure}(b). Across the
transition, $d_{\mathrm{Fe-Fe}}$ expands to 2.61\AA,
$d_{\mathrm{Fe-Fe}}'$ is further reduced to 2.57\AA, and $\beta$
jumps from 94.35$^{\circ}$ to 104.61$^{\circ}$. The lattice constant
$a$ expands from 8.17\AA\ to 8.32\AA, while $c$ is severely
compressed from 13.24\AA\ to 12.41\AA.  The difference in
$z_{\mathrm{Fe}}$ is enhanced from 0.07\AA\ to 0.26\AA. After the
transition, from 10 GPa to 14 GPa, $d_{\mathrm{Fe-Fe}}$ and
$d_{\mathrm{Fe-Fe}}'$ also slightly drops, but all Fe-Se-Fe angles
start to increase, and the difference in $z_{\mathrm{Fe}}$ keeps
increasing as well.

Another abrupt change in Fe-Fe distances (and Fe-Se-Fe angles) could
be identified from 14 GPa to 16 GPa (FIG.\ref{fig_ev_pressure}(b)),
as the lattice constant $a$ abnormally expands from 8.27\AA\  to
8.40\AA\, and $c$ collapses from 11.95\AA\ to 11.03\AA. A closer
examination reveals that the difference in $z_{\mathrm{Fe}}$ expands
abruptly from 0.32\AA\ at 14 GPa to 0.73\AA\ at 16 GPa. This severe
change implies that the backbone of vacancy superstructure is
becoming unstable and that the superstructure may break down under
such high pressure. Thus we focus on the magnetic phase transition
around 10 GPa in the following discussions.

\begin{figure}[ht]
 \centering
 \subfigure[$J$-$P$]{
   \rotatebox{270}{\scalebox{0.4}{\includegraphics{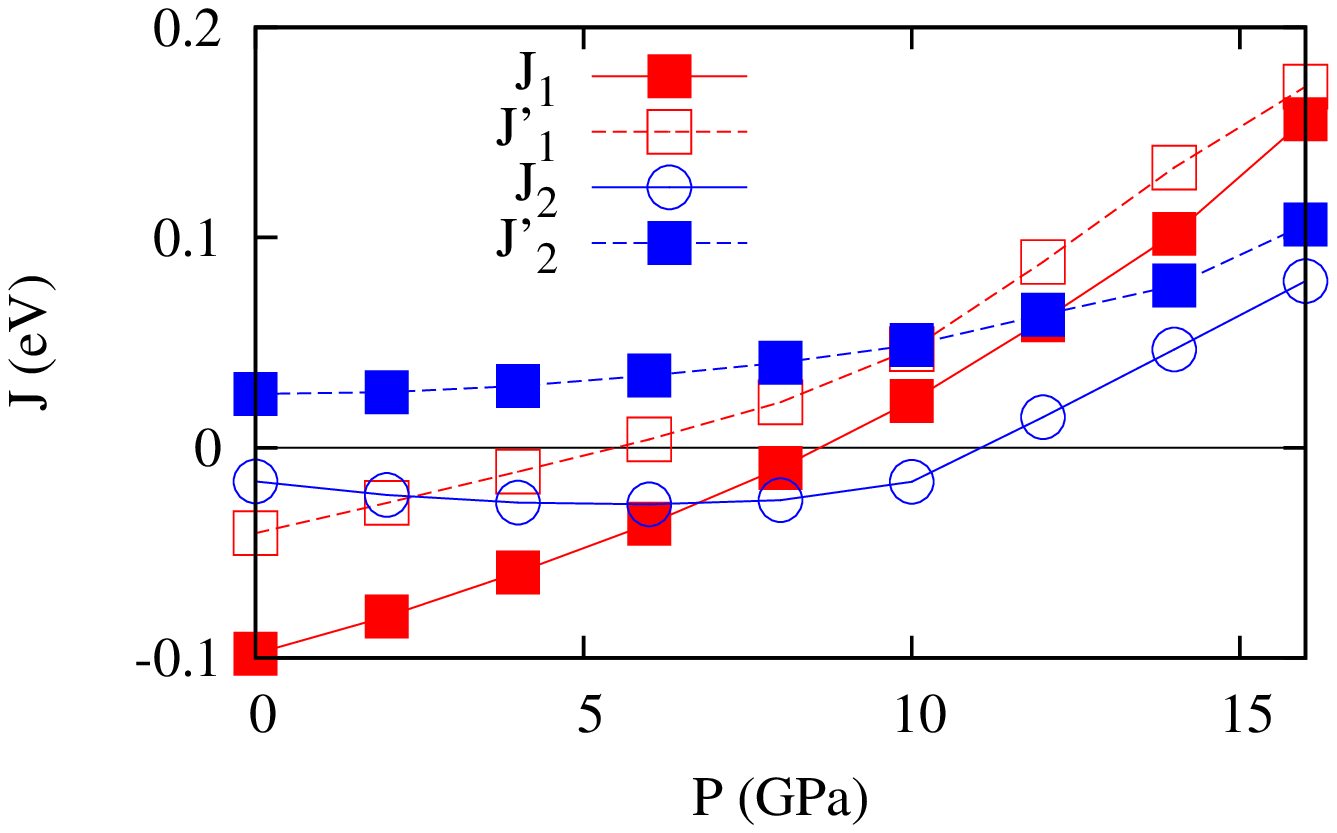}}}
   \label{fig_j_fit}
 }
 \subfigure[$E_g$ and $m_{\mathrm{Fe}}$]{
   \rotatebox{270}{\scalebox{0.4}{\includegraphics{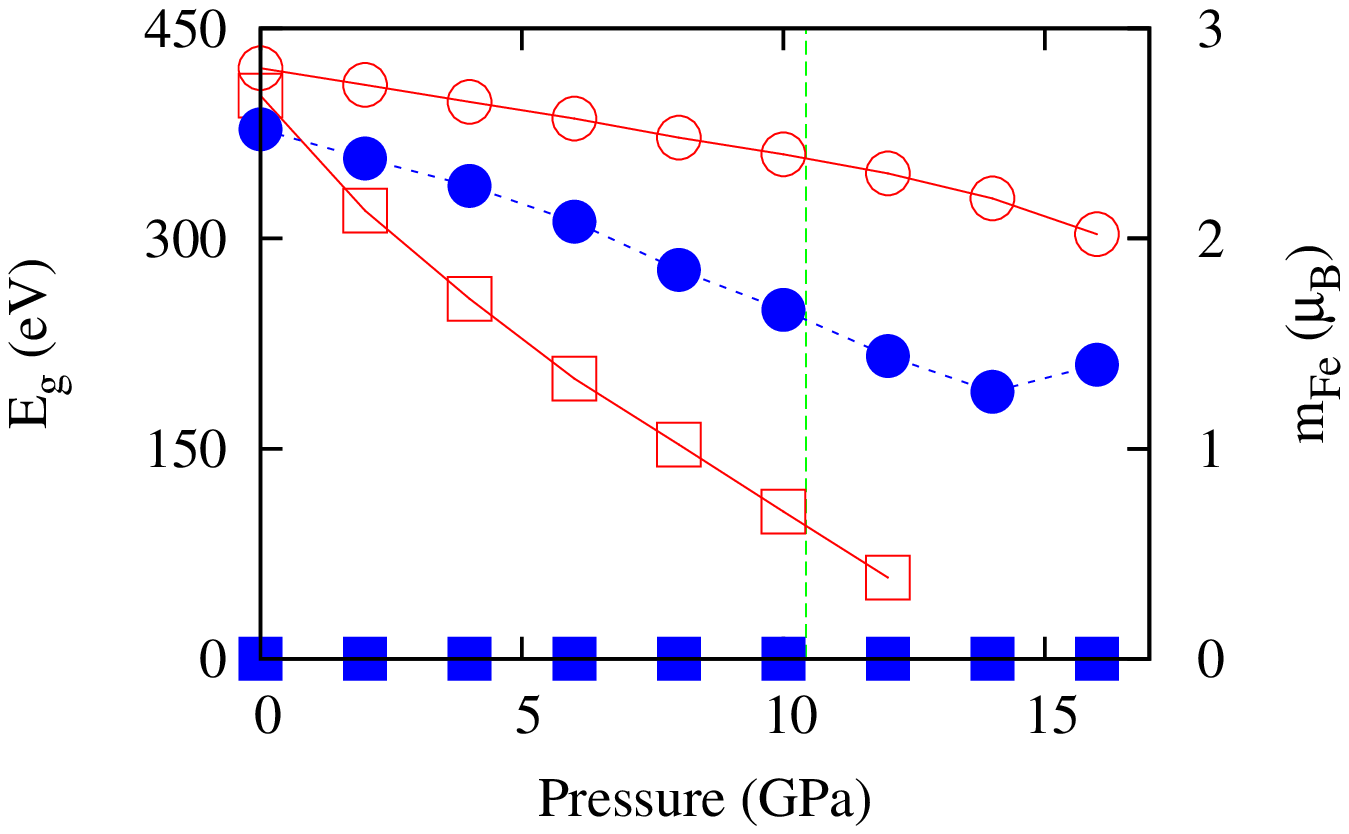}}}
   \label{fig_mgp}
 }
 \caption{ The variation of electronic structure with respect to external pressure. (a) Exchange-coupling constants fitted using the extended $J_1$-$J_2$ model, and (b) band gap size $E_g$ and magnetic moment per Fe $m_{\mathrm{Fe}}$. In panel (b), the open/filled circles and  squares denote $m_{\mathrm{Fe}}$ and $E_g$ at the BS-AFM/N\'{e}el-FM phases, respectively.\label{fig_estructure_p}}
\end{figure}

{\it Extended J$_1$-J$_2$ Heisenberg model.}
In order to understand the physics behind the pressure-induced
magnetic phase transition we fit the energetics of the magnetic
configurations using the extended $J_1$-$J_2$ model
\cite{cao_dai_245}(FIG. \ref{fig_estructure_p}(a)) where $J_1$ and
$J_2$ (or $J'_1$ and $J'_2$ ) are the intrablock (or the interblock)
n.n. and next-nearest-neighbor (n.n.n.) exchange couplings,
respectively. $J'_2$ is always AFM, for pressure from 0 GPa to 16
GPa; while $J_1$, $J'_1$ and $J_2$, being FM at 0 GPa, become AFM at
approximately 8 GPa, 6 GPa, and 11 GPa, respectively. All $J$s
increase monotonically with increasing pressure, except for $J_2$
which dwells around -20 meV until 10GPa. We notice that for ambient
pressure the fitted values of $J$s are compatible with the low
energy spin wave excitations probed in the recent
experiment\cite{arxiv:1105.4675}, while in the high pressure
N\'{e}el-FM phase, the fitted values are compatible with the Monte
Carlo result\cite{arxiv:1104.1445}.

Variations of these exchange interactions, while reflecting
complicated electronic structures that evolve with pressure, can be
attributed to a combined effect of the electron correlation ($U$),
the Hund's coupling ($J_H$), the crystal field splitting, as well as
various short-ranged hopping integrals at the microscopic
five-orbital Hubbard model level. As inferred from the previous DFT
study for the iron pnictides\cite{LuXiangPRB2008}, $J_H$ plays an
important role in mapping out the local magnetic interactions though
$U$ involved in GGA calculations may be not large. Some insights can
be gained if we adopt the intuitive results obtained from
perturbation or Hartree-Fock mean-field
theory\cite{PhysRevLett.101.076401,arxiv:1107.2279}. The n.n. and
n.n.n. exchange couplings are then contributed from two virtual
processes where Fe-3d electrons hop between two sites directly, or
via p-orbitals of the Se-atoms\cite{LuXiangPRB2008}. The
contribution from the direct exchange depends strongly on the
inter-atomic distances, while the contribution from the indirect
exchange depends strongly on both the inter-atomic distance and the
Fe-Se-Fe angle.

On one hand, both contributions decrease monotonically with Hund's
coupling and become FM at relatively large $J_H/U$. On the other
hand, for fixed $J_H/U$, the second contribution could be
ferromagnetic when the angle $\gamma=(\pi-\beta)/2$ formed by the
Fe-Se bond and the Fe plane is larger than a certain threshold value
varying in between $35^{\circ}$ $\sim$ $40^{\circ}$ approximately,
depending on the details of materials\cite{arxiv:1107.2279}. From
our structural analysis, the threshold values of $\gamma$ calculated
by the structure parameters for the corresponding fitted $J_1$,
$J'_1$, and $J_{2}$ are about $36^{\circ}\sim42^{\circ}$, which
reasonably fall into the FM region given by the mean-field
approximation\cite{arxiv:1107.2279}. Meanwhile, the $\gamma$ angle
associated with $J'_2$ is the smallest (about $35^{\circ}$ or less)
due to the structural distortion. It explains why $J_1$, $J'_1$, and
$J_{2}$ are initially FM and increase under pressure, while $J'_{2}$
is AFM even for ambient pressure. It is interesting to remark that
across the phase transition, $\beta$ angle abruptly changes from
94.35$^{\circ}$ to 104.61$^{\circ}$, leading to a sign change in the
second contribution, and thus $J_2$ becomes AFM around 10 GPa. From
10 GPa to 14 GPa, not only Fe-Fe distances are reduced, but the
Fe-Se-Fe angles (or $\gamma$)are also increased (reduced), thus all
$J$s rapidly increases.

Once the two competing phases, namely the BS-AFM and N{\'e}el-FM
phases, are identified as shown in FIG.\ref{fig_ev_pressure}(a), the
magnetic phase transition point should be at $J_1=J'_1/2$, when the
energy of the N\'{e}el-FM phase is equal to that of the BS-AFM
phase. Using this criterion and the least-square-fitting of the
$J$s, we determine the critical pressure to be 10.53GPa. Hence the
phase transition around 10 GPa is mainly driven by competition
between $J_1$ and $J'_1$, while the role of $J_2$ or $J'_2$ is to
stabilize the BS-AFM phase or the N{\'e}el-FM phases in the
respective low or high pressure regimes\cite{note2}.

{\it Band structure and density of states.} The transition from the
BS-AFM to N\'{e}el-FM phases drastically changes the electronic
structure of the material. It was shown that
$A_{0.8}$Fe$_{1.6}$Se$_2$ is an AFM insulator with the band gap
$\sim$ 400-600 meV \cite{cao_dai_245,PhysRevB.83.233205}. Here, the
gap size $E_g$ and the magnetic moment per Fe atom $m_{{\mathrm
Fe}}$ of Tl$_{0.8}$Fe$_{1.6}$Se$_2$ in the BS-AFM or the N\'{e}el-FM
states are calculated and shown in FIG. \ref{fig_estructure_p}(b).

In the BS-AFM phase, when the pressure increases from 0 to 10GPa,
the intra- and inter-block Fe-Fe n.n. distances greatly reduce,
which enhance the hopping between the n.n. Fe atoms, resulting in a
decrease of $E_g$ from $\sim$400 meV to $\sim$100 meV. Meanwhile,
$m_{\mathrm{Fe}}$ starts from 2.81 $\mu_B$ at 0GPa and gradually
reduces to 2.40 $\mu_B$ at 10GPa. The rapid suppression of $E_g$
compared with the small variation of $m_{\mathrm{Fe}}$ reveals
possible vacancy-enhanced Mott physics
\cite{cao_dai_234,PhysRevLett.106.186401,0295-5075-95-1-17003}.

By contrary, we find  that the N\'{e}el-FM phase is always metallic
as shown in FIG.3(b). We have also checked that the metallicity is
robust until $U\sim$ 4 eV by using GGA+$U$ calculations. Therefore,
for $y=0.8$, the magnetic phase transition is associated with an
insulator-metal phase transition. In the N\'{e}el-FM phase,
$m_{\mathrm{Fe}}$ sets off from 2.52 $\mu_B$ at 0GPa, and is reduced
to 1.66 $\mu_B$ at 10 GPa. It reaches the lowest value of 1.27
$\mu_B$ at 14GPa, where the collapsed phase transition occurs and
restores $m_{\mathrm{Fe}}$ to 1.40 $\mu_B$ at 16GPa. It is important
to notice that although the N\'{e}el block ensures vanishing net
moment in the classic ordering configuration, there is no overall
$\mathbf{q}$-vector ensuring the time-reversal symmetry, since the
blocks are ferromagnetically aligned.

{\it Discussions.} The magnetic states considered in this paper are
based on the classical Heisenberg model, and the $m_{\mathrm{Fe}}$,
which is about 2.5$\mu_B$-3.0$\mu_B$ in the BS-AFM phase, should be
considered as the static local magnetic moment. This is in agreement
with the neutron diffraction experiment where a large iron moment
3.31$\mu_B$/Fe (for $A$=K) was reported \cite{arxiv:1102.0830}.
Therefore quantum fluctuations should be suppressed in the ordered
phases due to the large magnetic moment. One expects that quantum
fluctuations play a larger role when the magnetic transition point
is approached, but the two magnetic phases should be robust in the
presence of the vacancy superstructure. The magnetic phase
transition is likely of first order because it is accompanied with a
sizable change in the lattice constants. An adequate account of
quantum fluctuations needed for a thorough understanding of the
transition calls for future studies.

Our results shed a new light in understanding the available high
pressure experiment\cite{arxiv:1101.0092} if the resistance hump
observed in most of the superconducting samples of \afsxy\ is
interpreted as due to a phase separation involving a Fe-vacancy
disordered superconducting phase and a Fe-vacancy ordered BS-AFM
insulating phase, respectively. With this interpretation, the
metallic phase in the high pressure regime is not necessarily due to
the charge transfer between two iron sites of different occupancies
as previously expected, but due to the N\'eel-FM phase which
respects the $\sqrt 5\times\sqrt 5$ vacancy ordering. Of course, the
accompanied structural distortion, in particular the difference in
$z_{\mathrm{Fe}}$, may complicate the comparison with experiments.

\begin{acknowledgments}
We would like to thank Elihu Abrahams and Qimiao Si for careful
reading of the manuscript as well as for useful discussions. This
work has been supported by the NSFC, the NSF of Zhejiang Province
(No.Z6110033), the 973 Project of the MOST. All calculations were
performed at the High Performance Computing Center of Hangzhou
Normal University College of Science.
\end{acknowledgments}


\begin{thebibliography}{99}
\bibitem{note2}
When the structural distortion is sufficiently small under certain
circumstances, another magnetic phase, the AFM7 phase (see SI)
formed by ferromagnetically coupled collinear-AFM spin blocks, could
compete with the N\'{e}el-FM phase and become the ground state.
\end{thebibliography}
\end{document}


\section{\label{sec:level1}Supplementary Information}
\setcounter{figure}{0}
\setcounter{table}{0}
\makeatletter
\renewcommand \thefigure{S-\@arabic\c@figure}
\renewcommand \thetable{S-\@Roman\c@table}
\renewcommand \theequation{S\@arabic\c@equation}

 \subsection{Definition of magnetic patterns}

\begin{figure*}[ht]
 \centering
  \subfigure[AFM0]{
    \scalebox{0.12}{\includegraphics{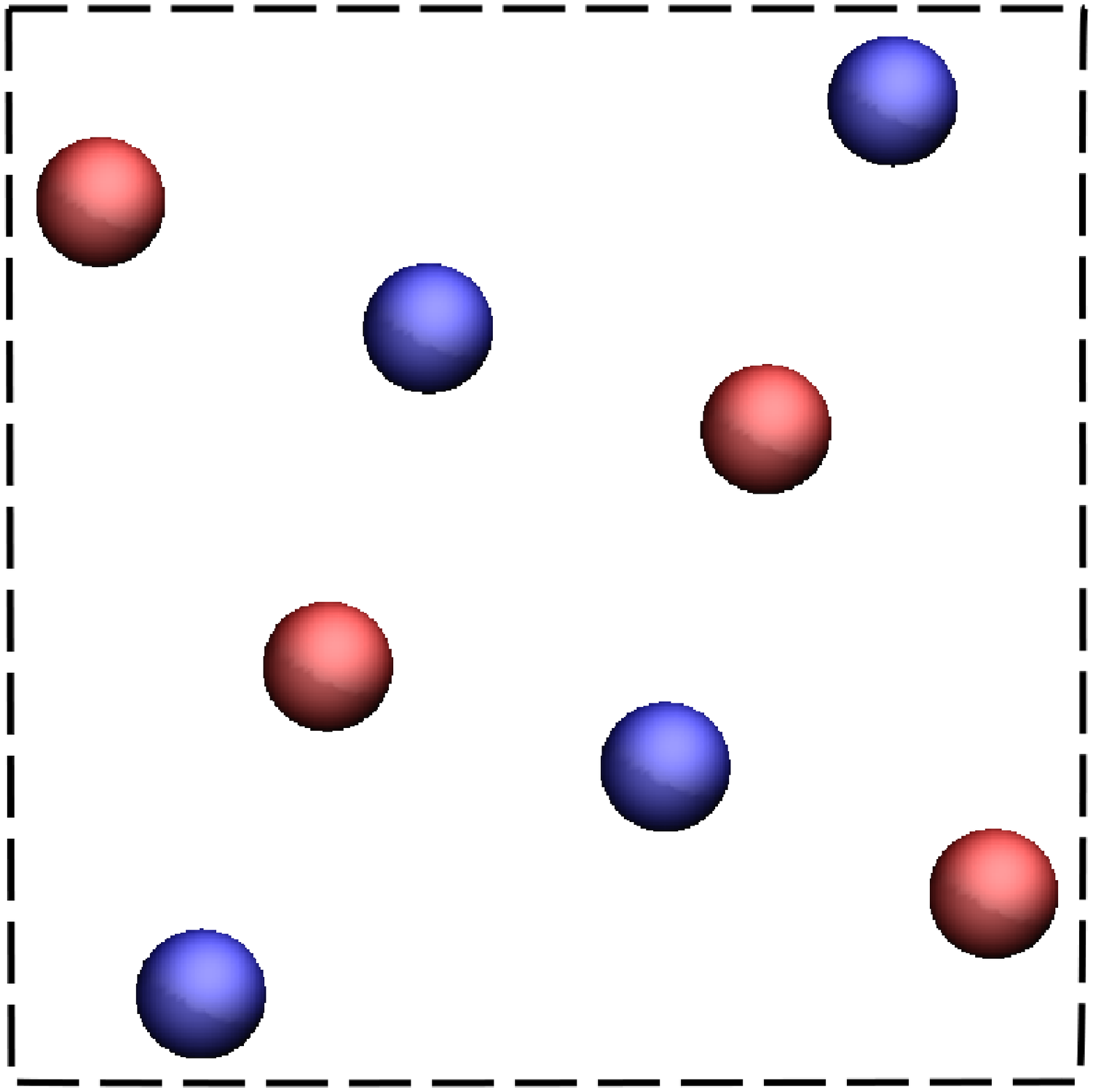}}}
  \subfigure[AFM1]{
    \scalebox{0.12}{\includegraphics{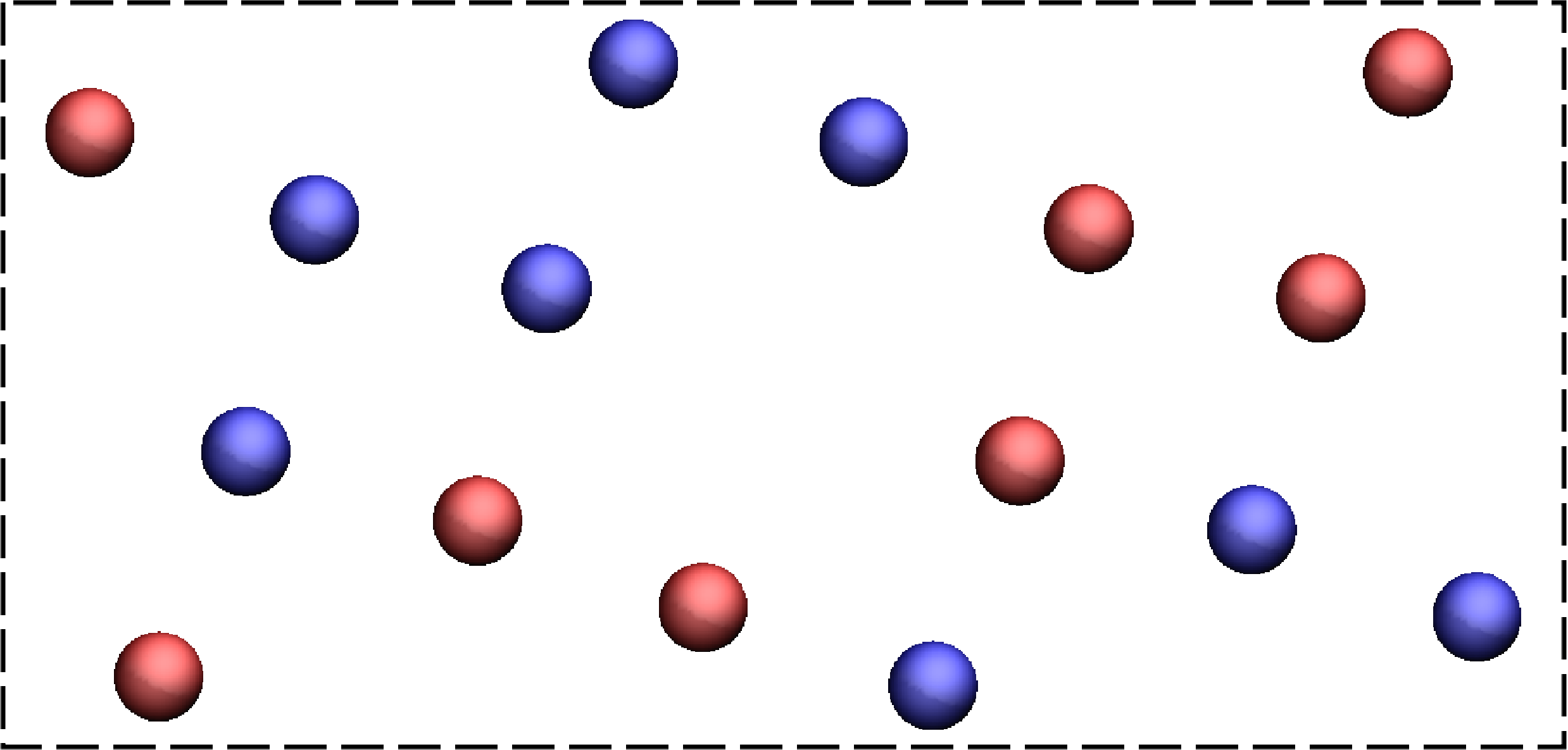}}}
  \subfigure[AFM2]{
    \scalebox{0.12}{\includegraphics{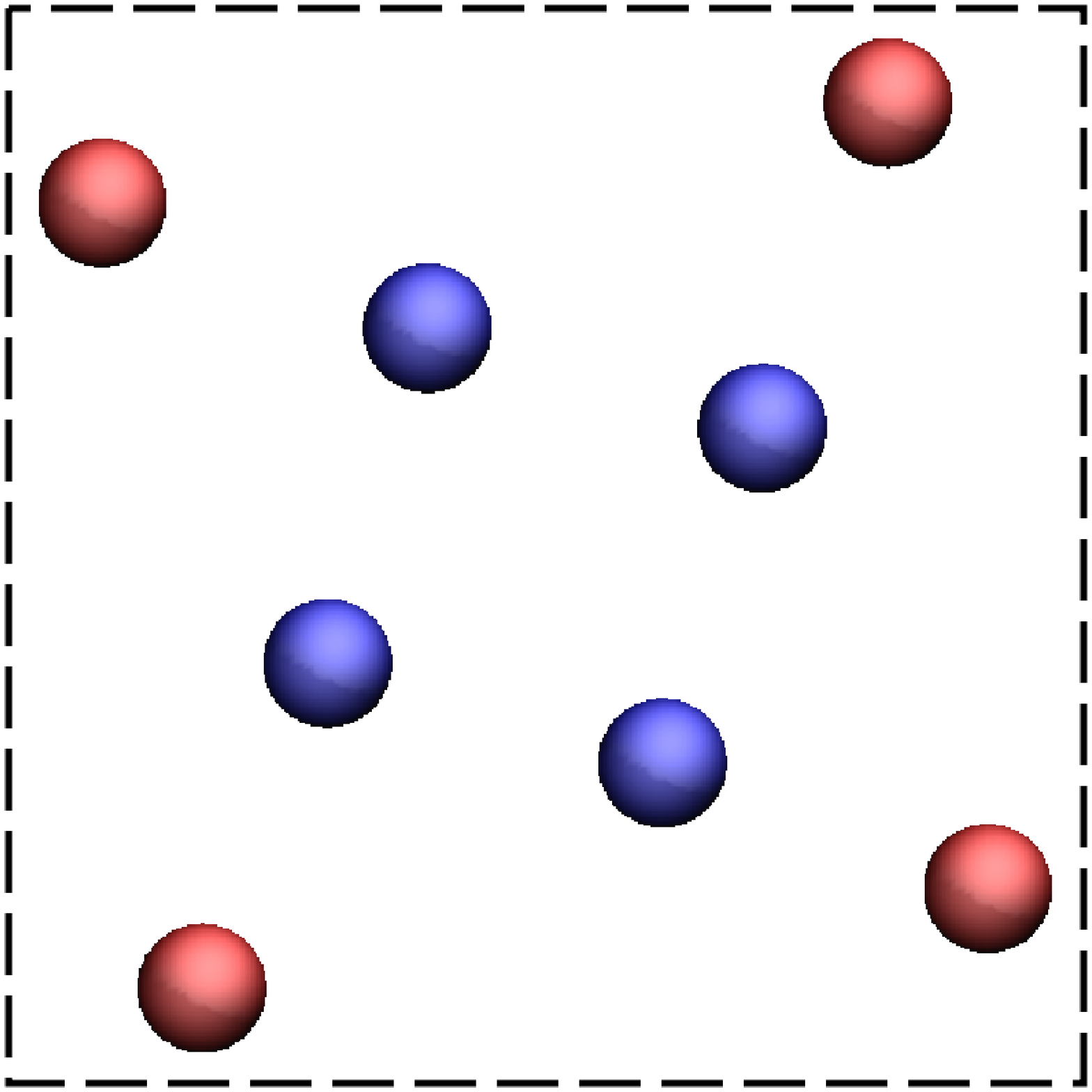}}}
  \newline
  \subfigure[AFM3]{
    \scalebox{0.12}{\includegraphics{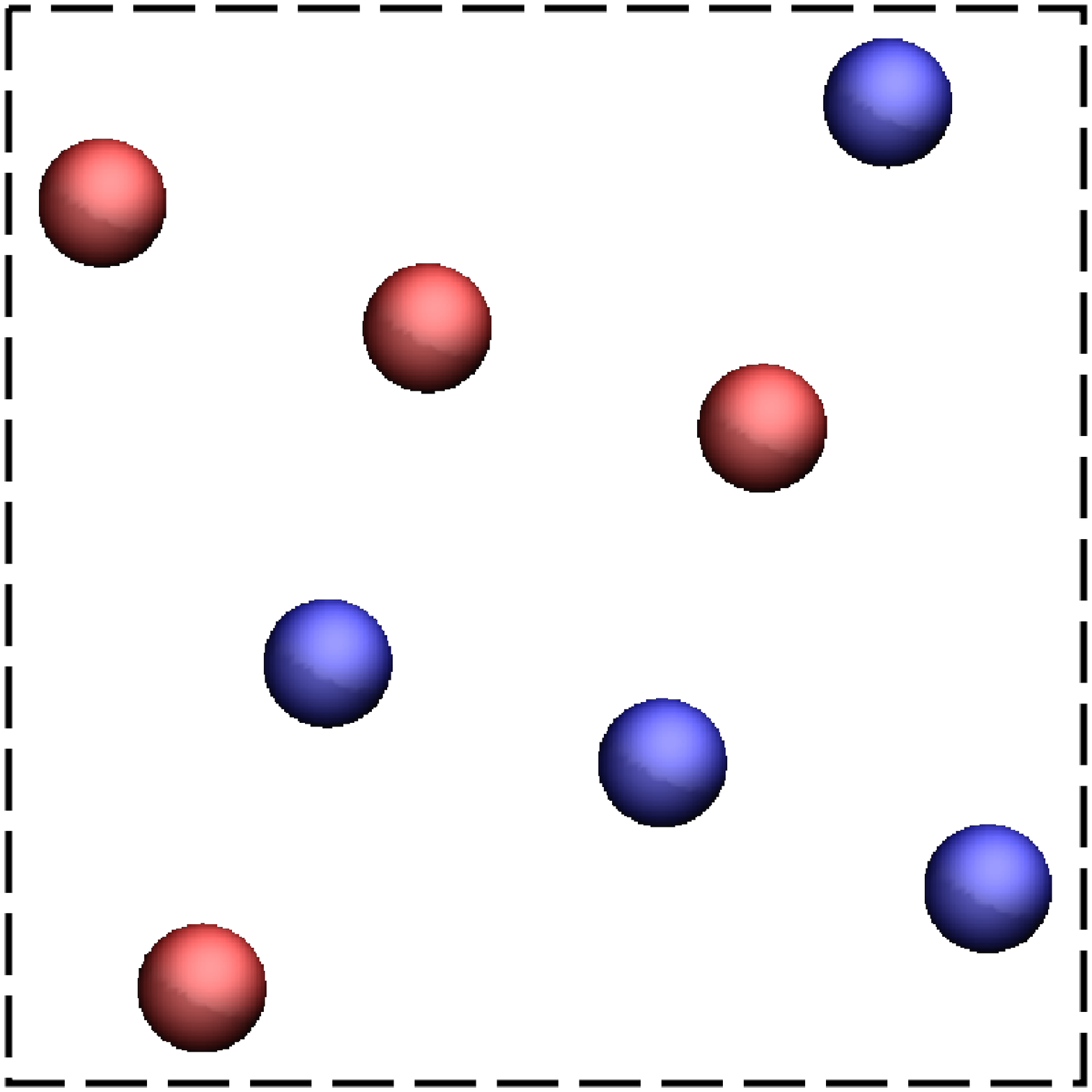}}}
  \subfigure[AFM4]{
    \scalebox{0.12}{\includegraphics{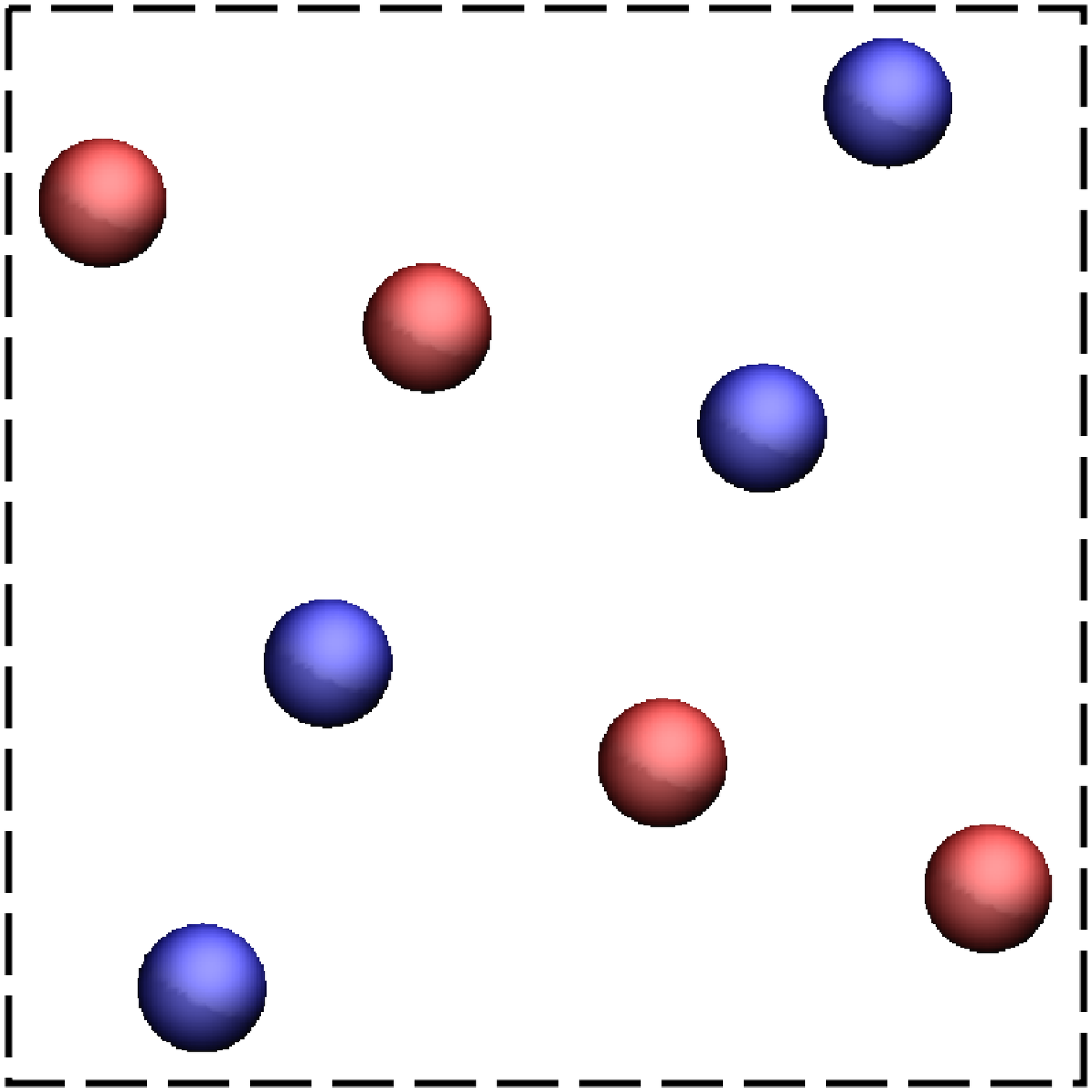}}}
  \subfigure[AFM5]{
    \scalebox{0.12}{\includegraphics{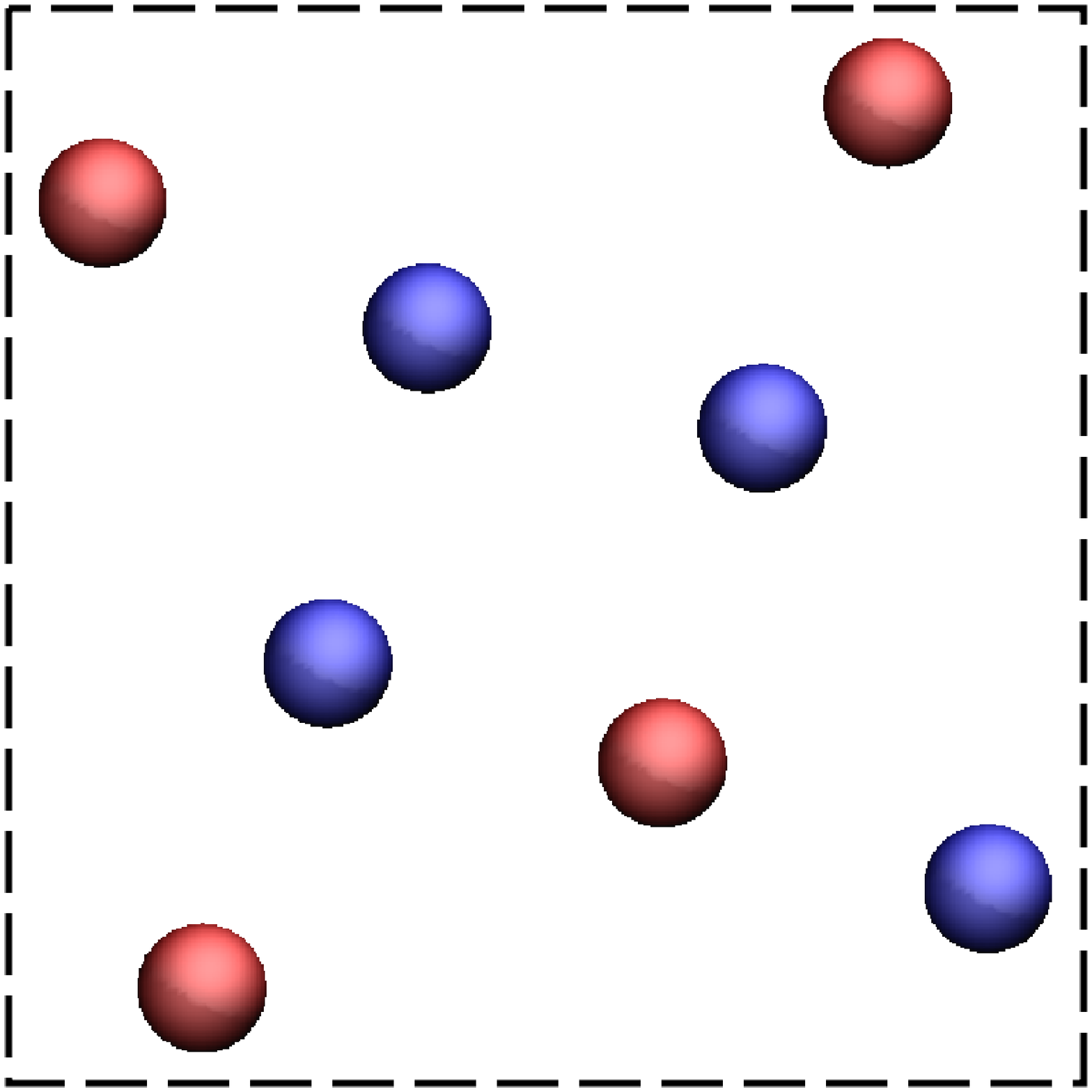}}}
  \subfigure[AFM6]{
    \scalebox{0.12}{\includegraphics{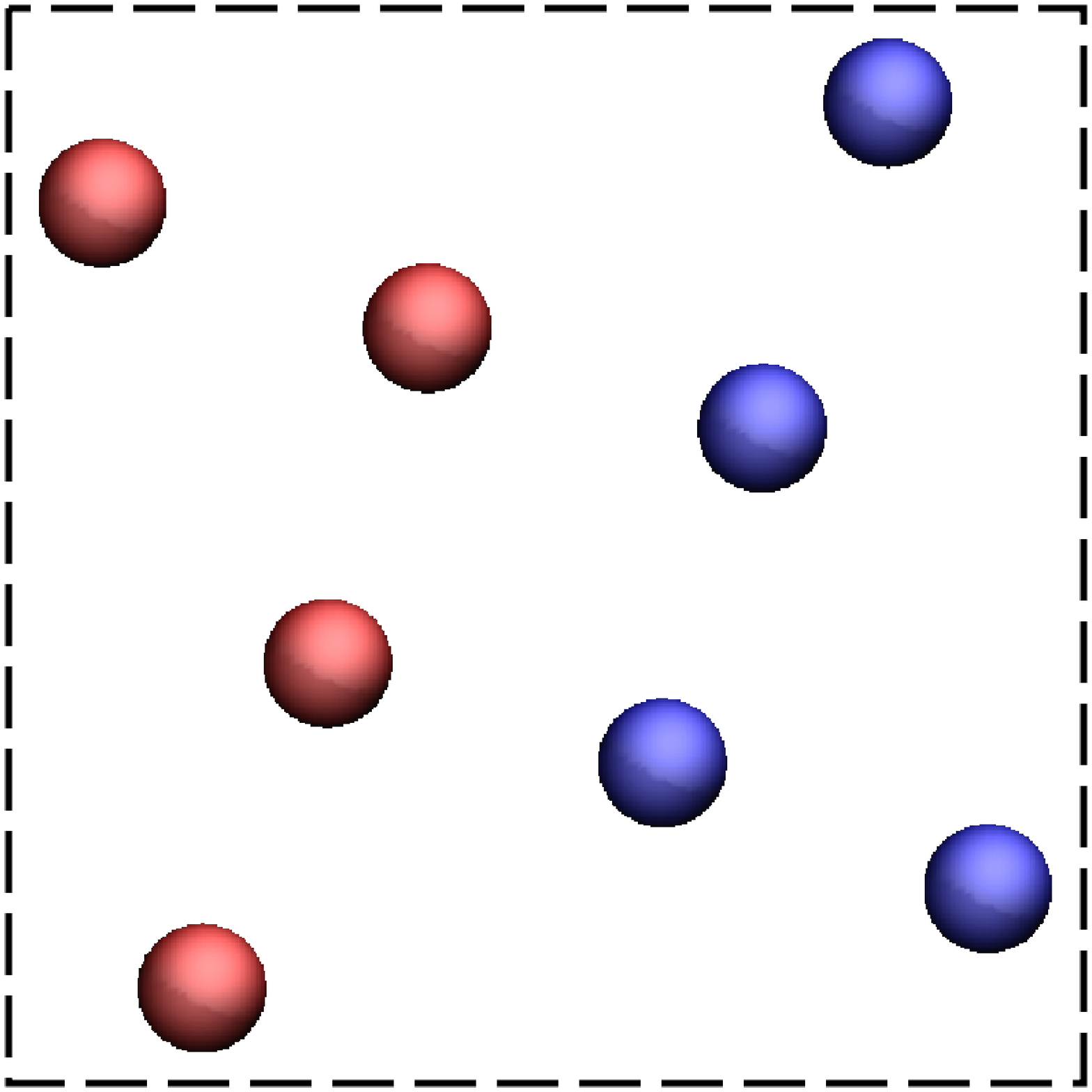}}}
  \subfigure[AFM7]{
    \scalebox{0.12}{\includegraphics{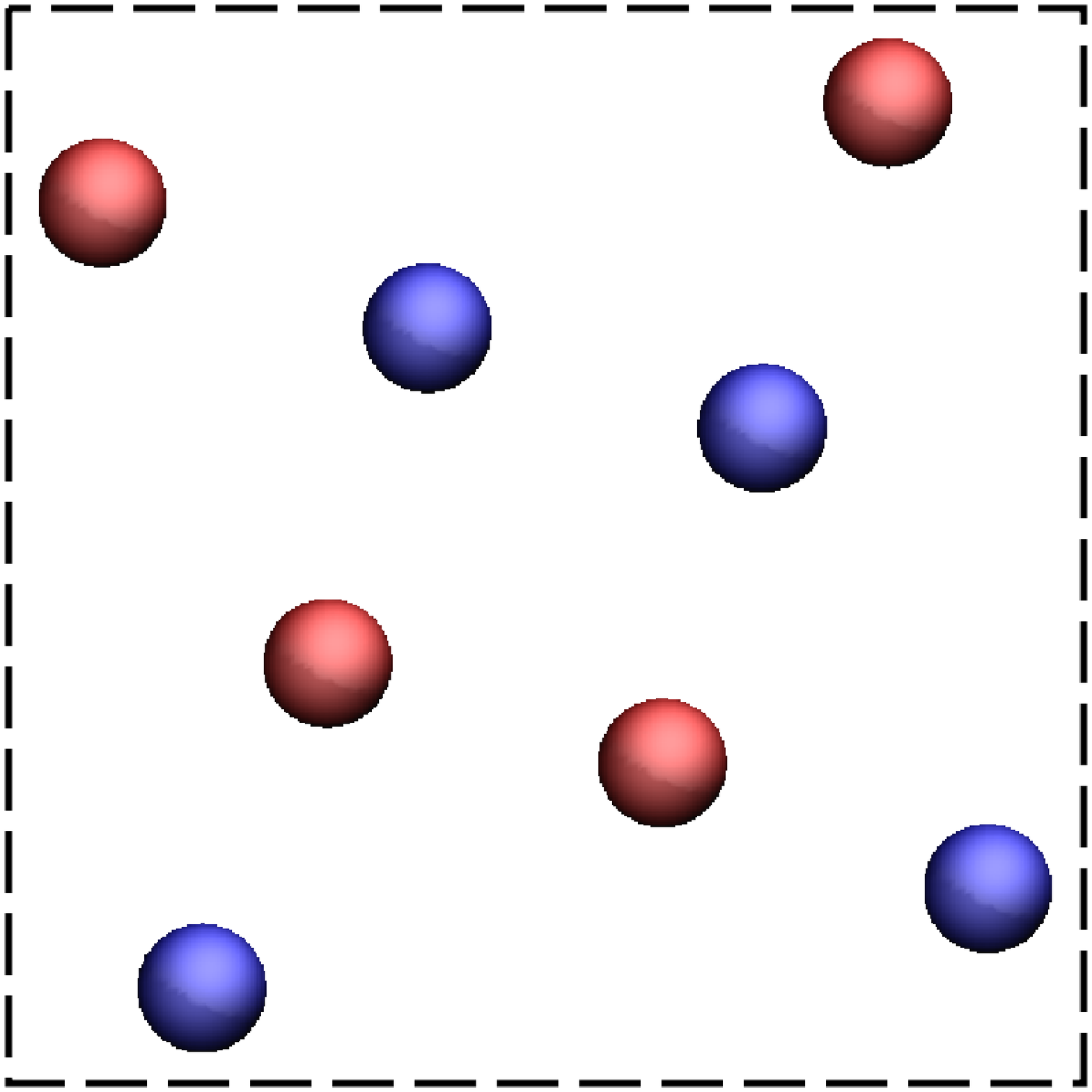}}}
  \caption{The magnetic structures considered.\label{SI_fig_1}}
\end{figure*}

We consider several possible ground states with magnetic
configurations or patterns as illustrated in FIG.(\ref{SI_fig_1}),
and we list their total energy under specific pressure in
TAB.(\ref{SI_tab_1}). These states can be captured by the extended
$J_1$-$J_2$ Heisenberg model in the classical limit
$$H=\sum_{<i,j>\mathrm{intra}}J_1 S_i S_j+\sum_{<i,j>\mathrm{inter}}J'_1 S_i S_j+\sum_{<<i,j>>\mathrm{intra}}J_2 S_i S_j+\sum_{<<i,j>>\mathrm{inter}}J'_2 S_i S_j. $$

\begin{table*}[h]
  \centering
  \caption{Total energies (in eV) per unit cell of different magnetic configurations at different pressures.}
  \begin{tabular}{c|c|c|c|c|c|c|c}
  \hline
  P(GPa) & AFM0 & AFM1 & AFM2 & AFM4 & AFM5 & AFM6 & AFM7 \\
  \hline
0 & -120.163 & -121.152 & -120.709 & -120.517 & -120.559 & -120.719 & -120.583 \\
2 & -113.772 & -114.625 & -114.201 & -114.076 & -114.094 & -114.175 & -114.094 \\
4 & -107.688 & -108.397 & -108.010 & -107.956 & -107.932 & -107.942 & -107.919 \\
6 & -101.841 & -102.405 & -102.061 & -102.088 & -102.013 & -101.956 & -101.994 \\
8 & -96.207 & -96.607 & -96.313 & -96.431 & -96.295 & -96.178 & -96.289 \\
10 & -90.767 & -90.980 & -90.750 & -90.955 & -90.766 & -90.581 & -90.798 \\
12 & -85.489 & -85.510 & -85.494 & -85.635 & -85.496 & -85.138 & -85.505 \\
14 & -80.366 & -80.169 & -80.364 & -80.450 & -80.364 & -79.834 & -80.366 \\
16 & -75.355 & -74.955 & -75.355 & -75.526 & -75.355 & -74.685 & -75.355 \\
  \hline
  \end{tabular}
  \label{SI_tab_1}
\end{table*}

\subsection{Energy expressions}
~The eigen energies of each states per block are listed as follows:
\begin{eqnarray}
  E_{\mathrm{AFM0}} &=& (-4J_1-2J'_1+2J_2+4J'_2)S^2 \label{SI_eqn_afm0}\\
  E_{\mathrm{AFM1}} &=& 0 \label{SI_eqn_afm1}\\
  E_{\mathrm{AFM2}} &=& (4J_1-2J'_1+2J_2-4J'_2)S^2 \label{SI_eqn_afm2}\\
  E_{\mathrm{AFM3}} &=& -2J_2S^2 \label{SI_eqn_afm3}\\
  E_{\mathrm{AFM4}} &=& (-4J_1+2J'_1+2J_2-4J'_2)S^2 \label{SI_eqn_afm4}\\
  E_{\mathrm{AFM5}} &=& -2J'_1S^2 \label{SI_eqn_afm5}\\
  E_{\mathrm{AFM6}} &=& (2J'_1-2J_2)S^2 \label{SI_eqn_afm6}\\
  E_{\mathrm{AFM7}} &=& (-2J'_1-2J_2)S^2 \label{SI_eqn_afm7}
\label{SI_eqn_1}
\end{eqnarray}

\subsection{Bi-collinear phase cannot be ground state}

In our discussions, the conventional FM phase as well as the
bi-collinear AFM phase (AFM1, Fig. \ref{SI_fig_1}(b)) observed in some 11-type iron
chalcogenides are excluded from the ground state candidates.

Firstly, we have performed simulations  at 12 GPa, confirming that
the bi-collinear and FM phases are about $\sim$ 13 meV/Fe and $\sim$
19 meV/Fe higher than the N\'{e}el-FM phase, respectively.

Secondly, based on the extended $J_1$-$J_2$ model, it is
straightforward to show that the bi-collinear phase could not be the
ground state when the $\sqrt{5}\times\sqrt{5}$ vacancy
superstructure is preserved.

{\it Proof:} If the bi-collinear phase (AFM1) were the ground state,
we could obtain from Eqn. \ref{SI_eqn_afm0} and Eqn.
\ref{SI_eqn_afm2}:  $-4J_1-2J'_1+2J_2+4J'_2>0$ and
$4J_1-2J'_1+2J_2-4J'_2>0$, respectively. Adding them together yields
$-J'_1+J_2>0$. However, this condition contradicts Eqn.
\ref{SI_eqn_afm6}, which yields $J'_1-J_2>0$. Therefore, the
bi-collinear phase could not be the ground state of the extended
$J_1$-$J_2$ model for any $J$s. The extra next-nearest-neighbor
exchanges $J_3$ and $J'_3$ are needed for this state to be the
ground state as in the conventional $J_1$-$J_2$-$J_3$ model.

  \subsection{Stability of vacancy superstructure at high pressure}
We have performed simulations using $5\times 5$ supercell (230 atoms
in all) with 3 randomly distributed Fe-vacancy patterns to verify
that the vacancy superstructure remains robust under high pressure
at least up to 12GPa. At 12GPa, the disordered vacancy
superstructures are at least 4 meV/Fe higher than non-magnetic
$\sqrt{5}\times\sqrt{5}$ phase, and are at least 23 meV/Fe higher
than the N\'{e}el-FM phase.

  \subsection{Critical point for the BS-AFM/N\'{e}el-FM phase transition}
~For large positive $J'_2$ or negative $J_2$, the ground state could
be either the BS-AFM state or the N{\'e}el-FM state. The competition
is given by
$E_{\mathrm{BS-AFM}}-E_{\mathrm{N\acute{e}el-FM}}=E_{\mathrm{AFM2}}-E_{\mathrm{AFM4}}=4S^2(2J_1-J'_1)$.
Hence the critical point is $J_1=J'_1/2$.

  \subsection{Band structure and DOS of the N\'{e}el-FM phase}

In the N\'{e}el-FM phase, the band structure indicates an electron
pocket for majority spin and a hole pocket for minority spin around
$\Gamma$. Two large electron pockets can also be identified for the
majority spin around $X$, where no structure exists for the
minority spin. If the occupancy of Tl sites is reduced by 20\%
(Tl$_{0.8}$Fe$_{1.6}$Se$_2$), the electron pocket around $\Gamma$
disappears due to the hole doping, and two hole pockets form around
$\Gamma$. In this case, one another hole pocket also shows up around
$X$. The electron states around $E_F$ are dominated by the Fe-3d
orbitals, while the Fe-3d and Se-4p orbitals hybridize over a wide
range from $E_F$-8 eV to $E_F$+2 eV.

\begin{figure}[ht]
 \centering
 \subfigure[Band Structure]{
   \rotatebox{270}{\scalebox{0.8}{\includegraphics{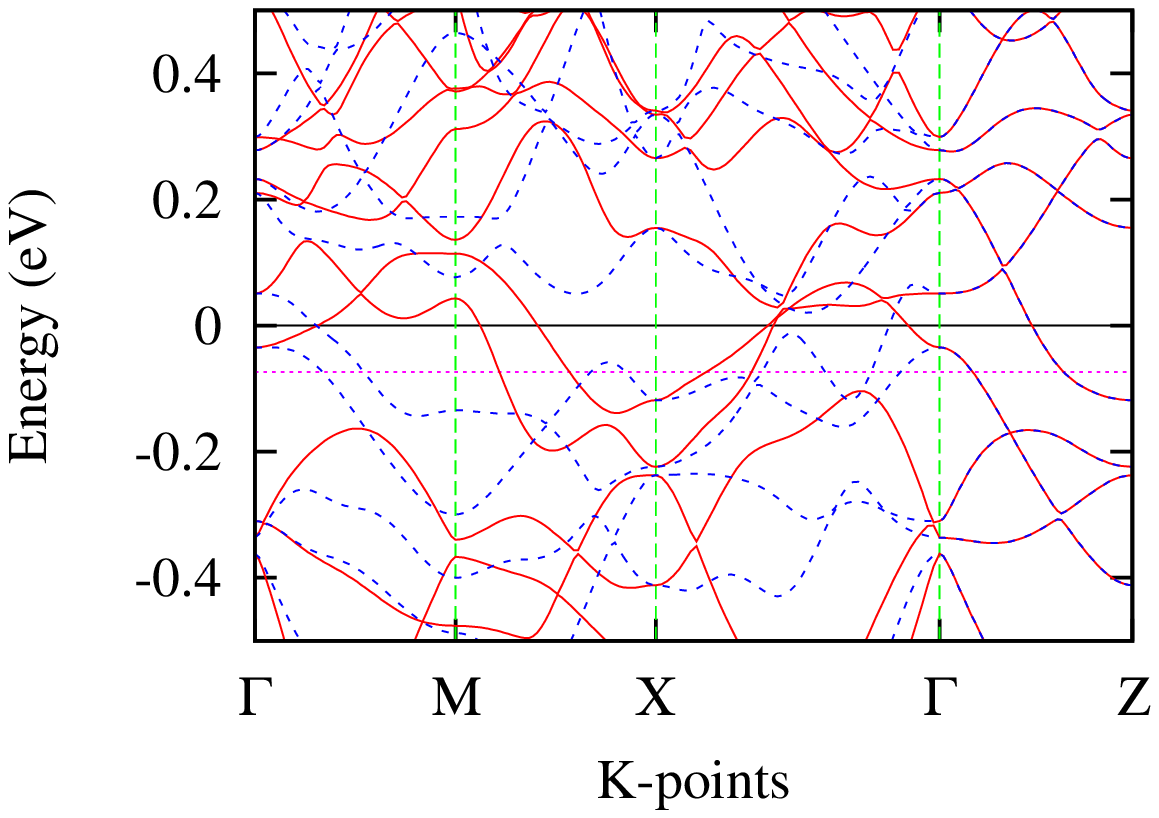}}}
   \label{fig_afm4_bs}
 }
 \subfigure[DOS]{
   \rotatebox{270}{\scalebox{0.8}{\includegraphics{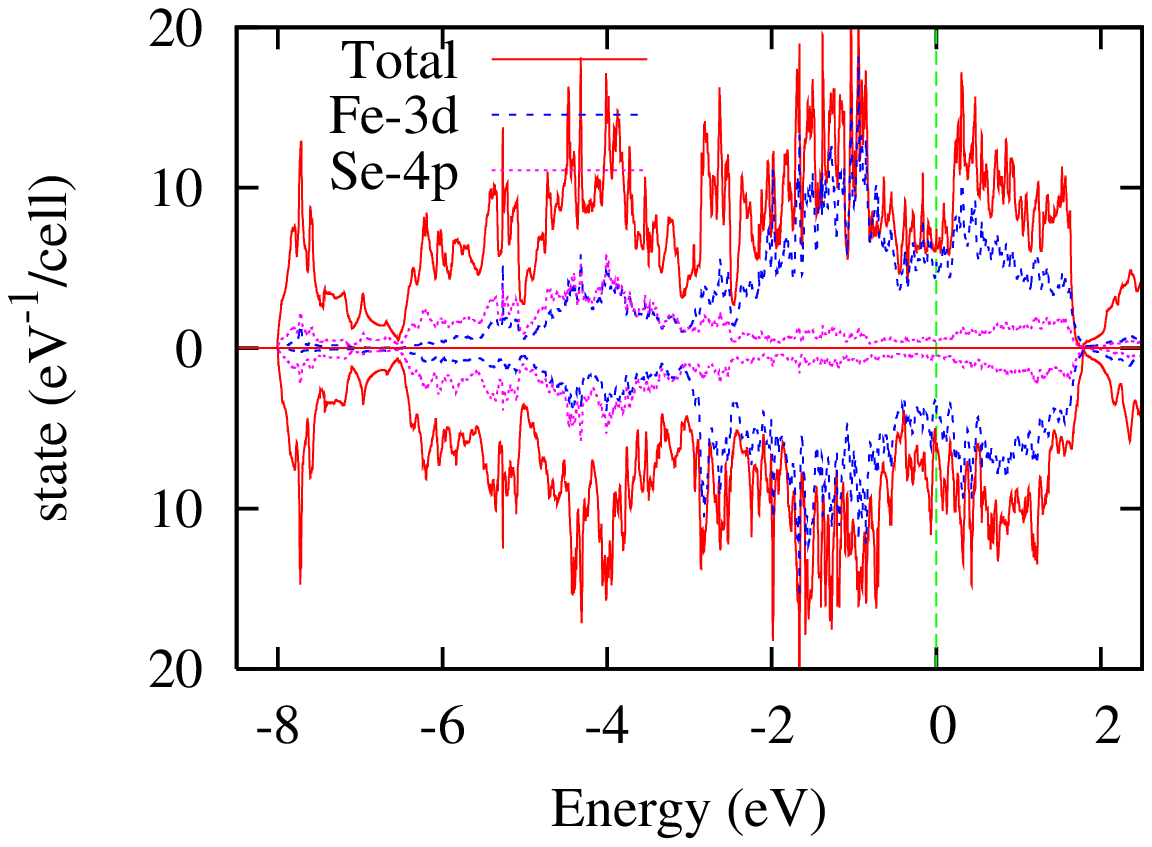}}}
   \label{fig_afm4_dos}
 }
 \caption{ a) Band structure and b) DOS of N\'{e}el-FM TlFe$_{1.6}$Se$_2$ at 12GPa. Notice that majority (solid red line in band structure; upper panel in DOS) and minority (dashed blue line in band structure; lower panel in DOS) spins are not degenerate in the $k_x$-$k_y$ plane, but are degenerate along $k_z$. \label{SI_fig_2}}
\end{figure}